\documentclass[10pt]{iopart}
\usepackage{iopams}
\usepackage{epsf}
\usepackage{times}
\usepackage{epsfig}
\usepackage{amsfonts,amsthm,eucal,amssymb}
\usepackage{color}
\usepackage{graphicx}
\eqnobysec
\begin{document}

\title{Carbon nanotubes in an inhomogeneous transverse magnetic field: exactly solvable model}
\author{
V Jakubsk\'y${}^*$, \c{S} Kuru${}^\dagger$ and J Negro${}^\ddagger$
}

\address{
{${}^*$}Department of Theoretical Physics,
Nuclear Physics Institute,
25068 Re\v z, Czech Republic\\
${}^\dagger$Department of Physics, Faculty of Sciences, Ankara
University, 06100 Ankara, Turkey\\
${}^\ddagger$Departamento de F\'{\i}sica Te\'orica, At\'omica y
\'Optica, Universidad de Valladolid, 47071 Valladolid, Spain }

\ead{jakub@ujf.cas.cz,kuru@science.ankara.edu.tr, jnegro@fta.uva.es}


\begin{abstract}

A class of exactly solvable models describing carbon nanotubes in the presence of an external inhomogeneous magnetic field is considered.
The framework of the continuum approximation is employed, where the motion of the charge carriers is governed by the Dirac-Weyl equation.
The explicit solution of a particular example is provided. It is  shown that these models possess nontrivial integrals of motion that
establish $N=2$ nonlinear supersymmetry in case of metallic and maximally semiconducting nanotubes. Remarkable stability of energy levels
with respect to small fluctuations of longitudinal momentum is demonstrated.
\end{abstract}

\section{Introduction}

Despite their structural simplicity, carbon nanotubes possess exciting physical properties \cite{Blase,NanotubesObserved}. Besides their 
ultimate strength \cite{UltimateStrength}  and elasticity \cite{Elasticity}, they particularly excel in the variability of their electronic 
characteristics that make them an attractive material for the applications in  electronic devices \cite{NanotubeElectronics,nttransistor}. 
Single-wall carbon nanotubes are small cylinders rolled up from
a graphene strip with the shell being just one atom-thick. The gap between their valence and conduction band (positive and negative energies) 
depends on the diameter of the nanotube and on the orientation of the lattice in the shell. Most of  nanotubes are semiconducting, they have 
a nonzero gap in the spectrum with magnitude proportional to the inverse of the diameter. The gap can be altered either by external fields 
or mechanical deformations.

The spectrum of carbon nanotubes in  the presence of external magnetic and electric field was discussed in numerous works with  the use of 
different techniques. For instance,   the band structure of the nanotubes was considered within the framework of the tight-binding Hamiltonian 
\cite{homogelectric,transverseelectric,transverseboth}. In \cite{ABoscillations}, it was shown that the metallic nanotubes can be turned into 
semi-conducting ones (and vice-versa) by an external homogeneous magnetic field, parallel with the axis of the nanotube. In \cite{Novikov}, 
the effect of a homogeneous transverse field on the spectrum was analyzed in the low-energy approximation.

In  this paper, we consider the single-wall carbon nanotube in the presence of a specific external magnetic field that is constant in the 
longitudinal direction (parallel with the axis of the nanotube), however, it has inhomogeneous transverse component (perpendicular to the axis). 
The setting is analyzed within the framework of the low-energy approximation. We show that dynamics of the spin-up and spin-down components is 
described by the associated Lam\'e equation.  The properties of the general finite-gap Hamiltonians are employed extensively in the analysis 
of this equation. 

The paper is organized as follows. In the next section, we briefly review the general description of single-wall carbon nanotubes in the presence 
of external magnetic field, explaining how the parallel and transverse fields affect the dynamics. In Section 3, we will introduce a solvable 
model that allows for explicit solutions of the stationary equation. Then, in Section 4, we show that the system possesses nontrivial symmetries 
that establish $N=2$ nonlinear supersymmetry for metallic and maximally semiconducting nanotubes. Next, in Section 5, we discuss the robustness 
of the energy levels with respect to a small fluctuation of the momentum in the longitudinal direction. The last section will be devoted to some 
comments and outlook.

\section{Nanotubes in the external magnetic field: the low energy approximation}
We focus on the spectral properties of single-wall carbon nanotubes in the low-energy regime where the  motion of quasi-particles is described 
by the Dirac-Weyl equation. Before considering the system in the external magnetic field, let us review briefly the case where the field is absent. 

\subsection{From graphene to nanotubes}
The electronic properties of nanotubes can be easily deduced from the characteristics of planar graphene. 
The tight-binding Hamiltonian of graphene in the vicinity of one of the Dirac points, where the energy goes to zero, reduces to the two-dimensional 
massless Dirac-Weyl Hamiltonian \cite{Semenoff}. The stationary equation in the $x$-$z$ plane can 
be written as \cite{Novikov} 
\begin{equation}\label{fse}
 H\Phi_{\epsilon}=\left(-i\sigma_2 \partial_z+i\sigma_1\partial_x\right)\Phi_{\epsilon}=\epsilon\,\Phi_{\epsilon},\qquad \epsilon=\frac{E}{v_F\hbar}\,.
\end{equation}
Here, $E$ is the energy and $v_F$ is the Fermi velocity that depends on the crystal characteristics (the hopping parameter) of graphene. The solution 
of (\ref{fse}) corresponding to a `scaled energy' $\epsilon$ is given by means of plane waves
\begin{equation}\label{fce}
\Phi_{\epsilon}(z,x)= e^{i\,(k_x x+k_z z)}\left(\begin{array}{c}
i \sqrt{i\,k_x - k_z}\\[1.5ex]
 \pm\sqrt{-i\,k_x - k_z}
\end{array}\right)
\end{equation}
where the upper (lower) sign corresponds to positive (negative) energies. Then, the dispersion relation  reads   
\begin{equation}\label{dr0}
\epsilon =  \pm\sqrt{k_x^2+  k_z^2}.
\end{equation}

The nanotube can be created by gluing together the long sides of a graphene strip. We fix the coordinates such that $x$ goes in the circumference 
direction, $0\leq x\leq 2\pi\rho_0$ with $\rho_0$ being the diameter, while $z$ is parallel to the axis of the nanotube. Specific boundary conditions 
have to be prescribed at $x=0$ and $x=2\pi\rho_0$. Their explicit form depends on the orientation of the hexagonal lattice in the strip; they are 
quasi-periodic in general \cite{Blase}. Thus, the solution of (\ref{fse})  has to satisfy the additional condition
\begin{equation}\label{boundarycondition}
\Phi_{\epsilon}(z,2\pi\rho_0)=e^{i \,2 \pi\omega}\Phi_{\epsilon}(z,0).
\end{equation}
The spectral properties of nanotubes, and hence their electronic properties, will depend on the periodic condition (\ref{boundarycondition}). As we 
shall see in the following, the phase factor $\omega$ is vanishing for metallic nanotubes whereas it acquires nonzero value for semi-conducting nanotubes.

\begin{figure}[h!]
\centering
\includegraphics[scale=0.5]{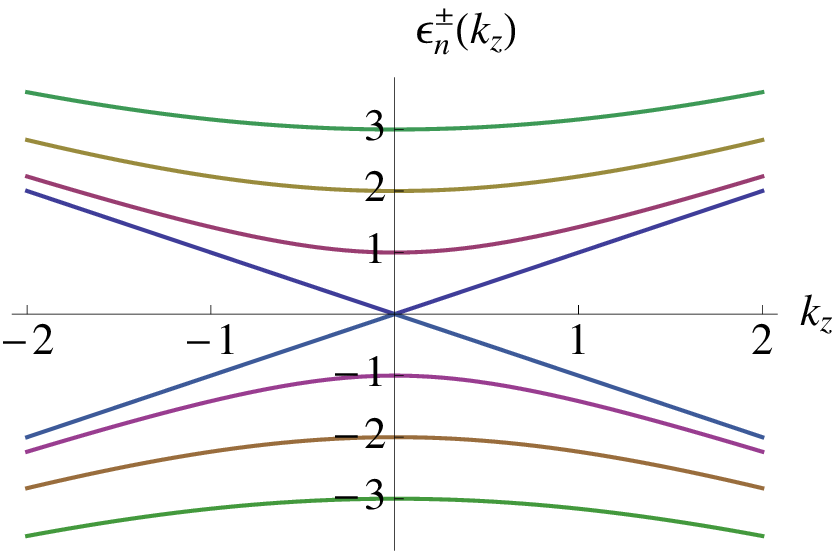}
\
\includegraphics[scale=0.5]{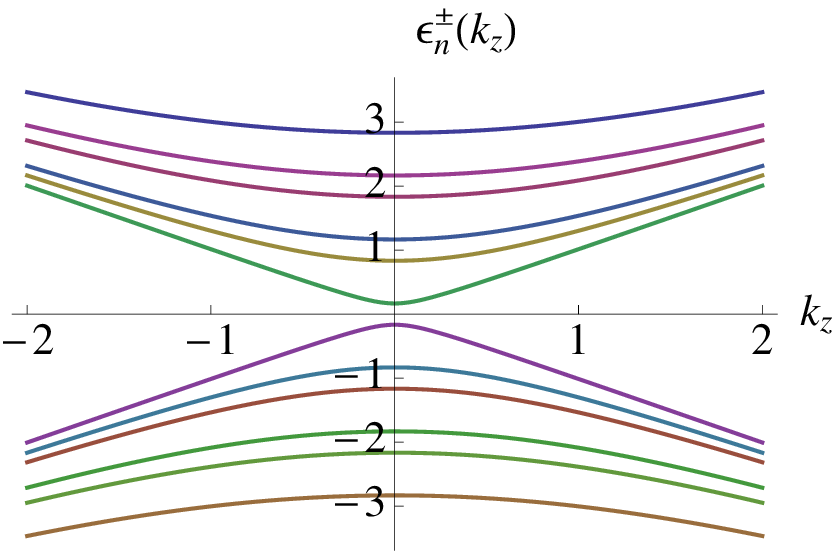}
\
\includegraphics[scale=0.5]{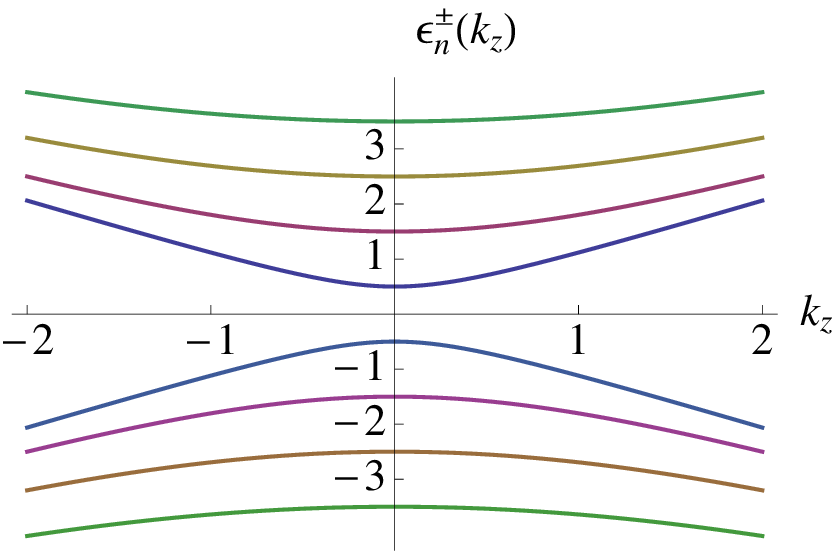}
\caption{The bands  (\ref{dr}) for metallic and semiconducting nanotubes
and some values of $n$: $\omega=0$ (left),
$\omega =1/6$ (center) and $\omega=1/2$ (right).
}\label{dispersions}
\end{figure}

In order to make the wave functions (\ref{fce}) compatible with the boundary condition (\ref{boundarycondition}), the momentum $k_x$ has to be 
restricted to the discrete values
\begin{equation}
k_x\equiv k_n = \frac{1}{\rho_0}(\omega+n), \qquad n \in \mathbb Z.
\end{equation}
The energies (\ref{dr0}) decay into a discrete set of values labeled  by $n$,  see Fig.\ \ref{dispersions} for illustration,
\begin{equation}\label{dr}
\epsilon_{ n}(k_z) =  \pm\sqrt{\frac{1}{\rho^2}(\omega +n)^2 +  k_z^2}.
\end{equation}
In particular, we have $\epsilon_{n}(0) = \pm\frac{1}{\rho_0}\,\left|\omega +n \right|$ for $k_z=0$.
The gap $\Delta(0)$ between  positive and  negative energies will depend on the explicit value of 
$\omega$,
\begin{equation}\label{dr2}
\Delta(0)=2|\epsilon_{0}(0)|=\frac{2}{\rho_0}|\omega| \, .
\end{equation}
For $\omega=0$, the gap is vanishing and the nanotubes are metallic since an infinitesimal excitation is sufficient to move the electron from valence 
to conduction band.  For $\omega\neq0$ a gap is opened as the minimal distance $\Delta(0)$ is nonzero.  The nanotube becomes semiconducting since an 
energy higher or equal to 
$\Delta(0)$ is needed to move the electron from negative to positive energy bands, which is the attribute of semiconductors. The formula (\ref{dr2}) 
implies that the gap is proportional to the inverse of the radius of the nanotube $\rho_0$.

\subsection{Nanotubes in external magnetic fields}

Now, let us consider the nanotubes in the presence of an external magnetic field. The magnetic field can be decomposed into transverse and 
longitudinal components (perpendicular and parallel to the axis of the nanotube). It induces the Lorentz force which is perpendicular to the 
movement of the charged particles. As the particles are confined to the two-dimensional surface, only the part of the transverse component 
which is normal to the surface has an effect on the trajectories of the particles. The longitudinal component  influences the dynamics as well, 
however, it manifests in a quite different manner. 

For convenience, we introduce $\phi=\frac{x}{\rho_0}$ where $\phi\in [0,2\pi]$. It constitutes, together with $z$ (axis of the nanotube) and 
$\rho_0$ (the radius), the cylindrical coordinates which parametrize the nanotube. The effective magnetic field $\bf B$ is induced by the vector 
potential tangent to the surface of the nanotube, ${\bf A} = A_\phi {\bf n}_\phi + A_z {\bf n}_z$, where ${\bf n}_\phi$ and ${\bf n}_z$ are the 
unit tangent vectors in the circumference and the longitudinal direction, respectively.  We will assume $A_{\phi}$ to be a constant and $A_z=A_z(\phi)$. 
The magnetic field is then \footnote{Notice that in  cylindrical coordinates, the curl of ${\bf A}={\bf n}_{\rho}A_{\rho}+ {\bf n}_{\phi}A_{\phi}+{\bf n}_{z}A_z$ 
reads 
\begin{equation}
\fl                                                                                                                                                       
{\bf \nabla}\times {\bf A}={\bf n}_{\rho}\left(\frac{1}{\rho}\partial_{\phi}A_z-\partial_zA_{\phi}\right)+{\bf n}_{\phi}\left(\partial_{z}A_{\rho}-\partial_{\rho}A_z\right)+{\bf n}_z\left(\partial_{\rho}A_{\phi}-\partial_{\phi}A_{\rho}+\frac{A_{\phi}}{\rho}\right).                                                                                                                                                                  \end{equation}
}
\begin{equation}\label{mgfield}
{\bf B} = \nabla \times {\bf A} = \frac1{\rho_0} \partial_\phi A_z(\phi) \,{\bf n}_\rho+ \frac{1}{\rho_0}A_{\phi}{\bf{n}}_z=B_{\perp}\,{\bf n}_\rho+B_{\parallel}{\bf{n}}_z.
\end{equation}
The massless Dirac-Weyl equation for the quasi-particle with minimal coupling to the magnetic field (\ref{mgfield}) takes the form 
\begin{equation}\label{r14c}
\fl
\left[-\sigma_2\left(i\partial_z+\frac{q}{c\,\hbar}A_z\right)+\sigma_1\left(\frac{i}{\rho_0}\partial_{\phi}
 +\frac{q}{c\,\hbar}A_\phi\right)\right]\hat\Phi(z,\phi)=\epsilon\,\hat \Phi(z,\phi)
 \end{equation}
where its solutions are subject to the boundary condition (\ref{boundarycondition}).
Taking into account the translational symmetry along the $z$ axis, we look for eigenfunctions
in the form $\hat\Phi(z,\phi) = e^{i\, k_z z}\hat\Phi(\phi) $ that, once replaced in (\ref{r14c}), gives
\begin{equation}\label{r14cc}
\fl
\left[-\sigma_2\left(- k_z+\frac{q}{c\,\hbar}A_z\right)+\sigma_1\left(\frac{i}{\rho_0}\partial_{\phi}
 +\frac{q}{c\,\hbar}A_\phi\right)\right]\hat\Phi(\phi)=\epsilon\,\hat \Phi(\phi)
\end{equation}

We can see that the effect of $A_{\phi}$ and $A_{z}$ on the dynamics of the quasi-particles is qualitatively different. The component $A_{\phi}$, 
corresponding to the longitudinal magnetic field, does not change the trajectory of a classical charged particle. However, as we will see now, 
it induces an additional phase shift of the wave functions, similarly to the Aharonov-Bohm effect. We can get rid of $A_{\phi}$ in the equation 
(\ref{r14c}) by extracting a suitable phase factor from the wave functions. Let us take 
\begin{equation}
 \hat\Phi(\phi)=e^{i\,\phi\frac{q\rho_0}{ch}A_{\phi}}\hat\Psi(\phi).
\end{equation}
Then we get the following equation for $\hat\Psi$,
\begin{equation}\label{r14}
\left[-\sigma_2\left(- k_z+\frac{q}{c\,\hbar}A_z\right)+\sigma_1\frac{i}{\rho_0}\partial_{\phi}\right]\hat\Psi(\phi)
=\epsilon\, \hat\Psi(\phi).
 \end{equation}
But now, the wave functions $\hat\Psi(\phi)$ are subject to the modified boundary condition
\begin{equation}\label{bcc}
 \hat\Psi(2\pi)=e^{i\,2\pi\delta}{\hat\Psi}(0),
 \qquad 
 \delta =\omega -\frac{q\rho_0}{c\hbar}A_{\phi}=\omega  -\frac{2S B_{\parallel}}{\Phi_0},
\end{equation}
where $S=\pi\rho_0^2$ is the section area of the nanotube and 
$\Phi_0=\frac{ch}{q}=\frac{2\pi c \hbar}{q}$. This shows that the phase $\delta$ depends on the magnetic flux $S B_{\parallel}$ that goes 
through the nanotube section.

In the current framework, we can easily understand the effect where the metallic nanotubes are converted into semi-conducting and vice-versa 
by the longitudinal magnetic field.  Fixing $A_{z}=0$, the energy gap of the nanotube can be obtained by substituting $\delta$ instead of 
$\omega$ into (\ref{dr2}).  Being interested in the lowest value of $|\epsilon_{n}(0)|=\frac{1}{\rho_0}\,\left|\delta +n \right|$,
we can consider $\delta\in [ 0,1/2]$ without  loss of generality  as the other values of $\delta$ can be compensated by the integer $n$. 
Now, the gap between the thresholds of the positive and negative energies becomes  
\begin{equation}\Delta(0)=\frac{2}{\rho_0}\left|\omega -\frac{q\rho_0}{c\hbar}A_{\phi}\right|.\end{equation}
By increasing $A_{\phi}$ adiabatically, the metallic nanotube ($\delta=0$) turns to be semiconducting. After reaching the phase where it 
is maximally semiconducting ($\delta=1/2$), the gap decreases   and the nanotube turns to be metallic  again.  This effect was called 
Aharonov-Bohm oscillation of carbon nanotubes \cite{ABoscillations}.

\section{Solvable model of almost homogeneous magnetic field: Finite-gap vector potential}
Let the carbon nanotube be immersed into the external magnetic field (\ref{mgfield}) where $B_{\parallel}$ is constant along the nanotube. 
The direction of the field can make an arbitrary angle with the axis of the nanotube. We will consider the configuration where
\begin{eqnarray}\label{approx}
A_z(\phi) &=& \rho_0 B_0\, (1+k')\,
 \frac{sn (\frac{(\phi+ {\pi}/{2})K}{\pi})\, cn(\frac{(\phi+{\pi}/{2})K}{\pi})}
 {dn(\frac{(\phi+{\pi}/{2})K}{\pi})}\nonumber\\
 &=&\rho_0 B_0a(\phi,k).
\end{eqnarray}
Here, $sn(x,k)$, $cn(x,k)$ and $dn(x,k)$ are Jacobi elliptic functions (the modulus $k$ has been suppressed in (\ref{approx}) to simplify the notation) 
and $K$ is for the complete elliptic integral
$K(k)=\int_0^{\frac{\pi}{2}}(1-k^2\sin^2 t)^{-1}dt$. The modular parameter is  $k\in [0,1]$, the complementary modulus will be $k'\equiv\sqrt{1-k^2}$ 
and  the notation $K\equiv K(k)$ or $K'\equiv
K(k')$ will be used. 

The vector potential (\ref{approx}) possesses some remarkable properties:

\begin{itemize}
\item[(a)] For $k_z=0$ and specific values of the intensity of the transverse field,
\begin{equation}\label{b0}
 B_0=m\frac{k^2Kc\hbar}{(1+k')\pi\rho_0^2\,q}
 =\frac{m k^2 K}{2\pi(1+k')}\frac{\Phi_0}{S}, \quad \Phi_0=\frac{ch}{q}=\frac{2\pi c \hbar}{q}
\end{equation}
the stationary equation (\ref{r14cc}) acquires the following form
\begin{equation}\label{DiracLame}
 H\Psi=
 \left(i\sigma_1\partial_y- m\,k^2 \,\frac{sn(y,k)cn(y,k)}{dn(y,k)}\,\sigma_2\right)\Psi
 =\tilde \epsilon\,\Psi
\end{equation}
where we have introduced the following notation
\begin{equation}\label{epsilon}
\fl
 y=\frac{K(\phi+{\pi}/{2})}{\pi},\quad
  \tilde \epsilon = \frac{\pi}{K}\,\rho_0\epsilon =
\frac{\pi}{K}\,\frac{\rho_0 }{v_F\hbar}\,E,\quad
\hat\Psi(\phi) = \Psi (y) \,.
\end{equation}
The Hamiltonian (\ref{DiracLame}) is $2K$ periodic  in the new variable  $y$. The solutions of the equation are subject to the boundary 
condition (\ref{bcc}), where the phase factor $\delta$ will also depend on $B_{\parallel}$, see (\ref{mgfield}), (\ref{bcc}). Thus, in 
the new variables, we have
\begin{equation}\label{bc2}
 \Psi(y+2K)=e^{i\,2\pi \delta}\Psi(y), \qquad 0\leq\delta\leq1/2.
\end{equation}

The key point is that when $m$ is an \textit{integer},  the Hamiltonian in (\ref{DiracLame})
can be classified as a {\it finite-gap} operator \cite{BdG,Gesztesy} and its (formal) eigenfunctions can be found analytically in terms 
of the Jacobi (or Weirstrass) theta and zeta functions, see \cite{Ganguly}. We will suppose that this is the case from now on.

\item[(b)] The  potential function (\ref{approx}) approximates very well to an homogenous field which is given by $A_{\phi}=\rho_0 B_0\cos\phi$, 
	see \cite{Novikov}. The smaller is $k$, the smaller is the difference between $\cos\phi$ and $a(\phi,k)$,
\begin{equation}
 \lim_{k\rightarrow0} a(\phi,k)=\cos \phi.
\end{equation}
Even for $k\sim 0.7$, the difference is of order $10^{-3}$, see Fig.\ \ref{fig2} for illustration.
Hence, the vector potential (\ref{approx}) gives rise to a magnetic field which fluctuates slightly around a constant value.
\end{itemize}
\newsavebox{\field}
    \savebox{\field}{
    \scalebox{1}{
    \includegraphics[scale=1]{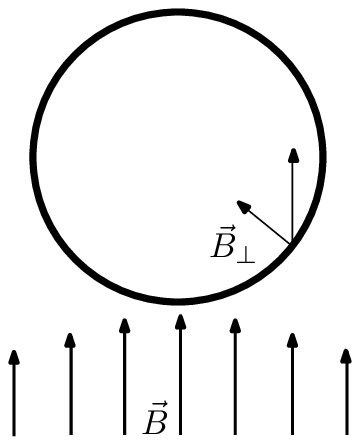}
    }
    }
\newsavebox{\difference}
    \savebox{\difference}{
    \scalebox{1}{
    \includegraphics[scale=.81]{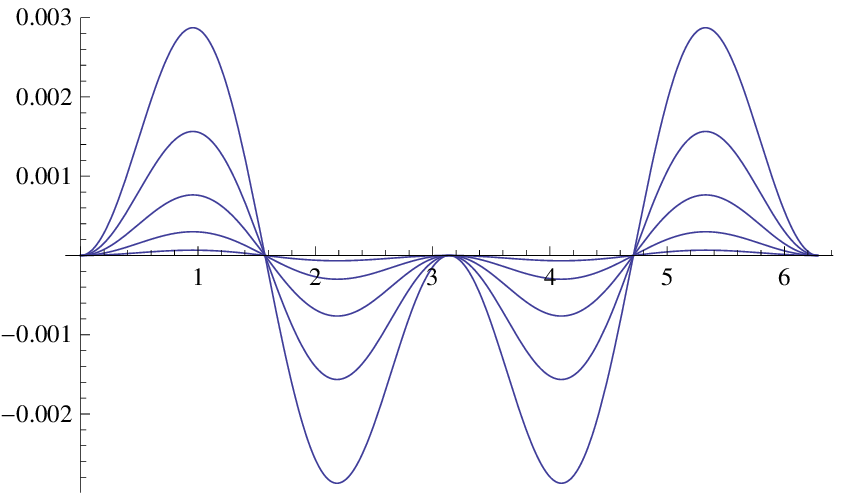}
    }
    }
\begin{figure}[h!]
\centering
\usebox{\field}\hspace{16mm}\usebox{\difference}
\caption{Left: Illustration of the transversal field $\vec{B}$ and its projection $\vec{B}_{\perp}$
to the normal of the surface of the nanotube. Right: The sum $a(\phi,k)-\cos(\phi)$ for $k^2\in\{0.5,0.4,0.3,0.2,0.1\}$. The lower value of $k$ the smaller is the sum.
}\label{fig2}
\end{figure}
Equation (\ref{DiracLame}) can be cast in the form
\begin{equation}\label{DiracLame2}
H\Psi= \left(\begin{array}{cc}
0 & i {\cal A}^\dagger
\\
-i {\cal A} & 0
\end{array}\right) \Psi = \tilde \epsilon \Psi
\end{equation}
where
\begin{equation}\label{aas}
\fl
\mathcal{A}^{\dagger}=\partial_y+mk^2\,\frac{sn(y,k)cn(y,k)}{dn(y,k)},
\quad
\mathcal{A}=-\partial_y+mk^2\,\frac{sn(y,k)cn(y,k)}{dn(y,k)} \,
\end{equation}
and $m$ is a fixed integer.
Then, the stationary equation (\ref{DiracLame}) can be diagonalized, 
\begin{eqnarray}\label{asocLame}
\fl H^2\, \Psi(y)\equiv\left(\begin{array}{cc}H_1&0\\
0& H_2\end{array}\right)\Psi(y) = 
\left(\begin{array}{cc}\mathcal{A}^{\dagger}\mathcal{A} &0\\
0& \mathcal{A}\mathcal{A}^{\dagger}\end{array}\right)\Psi(y)
&&
\nonumber\\
\fl =\left(\begin{array}{l}-\partial_{y}^2+ k^2 C_m^-\, sn^2(y,k)
+k^2C_m^+\,\frac{cn^2(y,k)} {dn^2(y,k)} - k^2 m^2 \qquad 0 \quad \\[1.5ex]
\quad 0\qquad -\partial_{y}^2 + k^2 C_m^+\, sn^2(y,k) + k^2C_m^-
\,\frac{cn^2(y,k)} {dn^2(y,k)} -k^2m^2\end{array}\right) \Psi(y) = \tilde \epsilon^2\Psi(y)
\end{eqnarray}
where $C_m^{\pm}=m(m\pm 1)$
\footnote{
The operator $H^2$ can be considered as the supersymmetric Hamiltonian that, together with the supercharges 
$Q^{\dagger}\equiv\frac{1}{2}(\sigma_1-i\sigma_2)\mathcal{A}^{\dagger}$ and
$Q\equiv\frac{1}{2}(\sigma_1+i\sigma_2)\mathcal{A}$, establishes the $sl(1|1)$ superalgebra \cite{cooper}.  
Denoting $H_{SUSY}\equiv H^2$, we have
\begin{eqnarray}
 && [H_{SUSY},Q^{\dagger}]=[H_{SUSY},Q]=0,\quad \{Q,Q^{\dagger}\}=2H_{SUSY},\nonumber
 \\ [1.5ex]
 && \{Q,Q\}=\{Q^{\dagger},Q^{\dagger}\}=0.
\end{eqnarray}
The framework of the SUSY quantum mechanics was utilized recently in the analysis of a  Dirac-Weyl system confined on 
the surface of the sphere \cite{cond13}.
}. 
The corresponding second-order equations for the spin up and down components of $\Psi(y)$ can be identified as the 
associated Lam\'e equations. It is worth noticing that the operators $H_1$ and $H_2$ are related by means of a shift 
of the coordinate in half of the period $K$: $H_2(y) = H_1(y+K)$.

Let us notice that the quantum systems described by the stationary equations (\ref{DiracLame}) and (\ref{asocLame}) 
with $y$ extending over the whole {\it real line} ($y\in\mathbb{R}$) and with integer valued $m$  were considered in 
the literature, see e.g.\ \cite{BdG,trisusy} and references therein. In these systems, the boundary condition (\ref{bc2})   
is replaced by the requirement that the wave functions be of Bloch type, i.e. $\delta$ can acquire any real value. Then, 
the spectrum consists of a finite number of spectral bands where the inner-band energy levels are doubly degenerate while 
the band-edge states are singlets. In our current case where the system is bounded on a finite interval, the boundary condition 
(\ref{bc2}) with fixed $\delta$ gives rise to a discrete spectrum. For instance, when $\delta=1/2$, just the anti-periodic 
states comply with (\ref{bc2}).

The finite-gap system described by (\ref{DiracLame}) possesses other interesting properties, e.g. the existence of a nontrivial 
integral of motion or the stability of the energy levels under the perturbation of small values of $k_z$, that will be discussed 
later in Sect.~5 and Sect.~6. Now, let us step to the explicit solution of the stationary equation (\ref{DiracLame}) for $m=1$.

\subsection*{Solutions for the configuration $m=1$ and $k_z=0$.}
Fixing $m=1$ in the field intensity (\ref{b0}),
the stationary equation (\ref{DiracLame}) acquires the following form
\begin{equation}\label{DiracLame2}
H\Psi(y)= 
\left(i\sigma_1\partial_y
 - k^2 \frac{sn(y,k)cn(y,k)}{dn(y,k)}\, \sigma_2\right)\Psi(y)=\tilde \epsilon\,\Psi(y).
\end{equation}
In this case, the diagonal operators of the decoupled equation in (\ref{asocLame})  take the simple form of Lam\'e equations
\begin{equation}\label{asocLame2}
\left(\begin{array}{l}-\partial_{y}^2
+  2 k^2 \, sn^2(y+K,k) - k^2  \qquad 0 \  \\[1.5ex]
 \quad 0\qquad \quad-\partial_{y}^2 + 2 k^2 \, sn^2(y,k)
 -k^2\end{array}\right)
\Psi(y) = \tilde\epsilon^2\Psi(y).
\end{equation}
To find $\Psi$, it is sufficient to solve just one of the Lam\'e equations in (\ref{asocLame2}). Thus, for instance, take 
that of spin down component
\begin{equation}\label{al}
H_2 \psi = \left(-\partial_{y}^2 + 2 k^2 \, sn^2(y,k)-k^2\right)\psi
 = \tilde\epsilon^2 \psi.
\end{equation}
\begin{figure}[h!]
\centering
\includegraphics[scale=0.8]{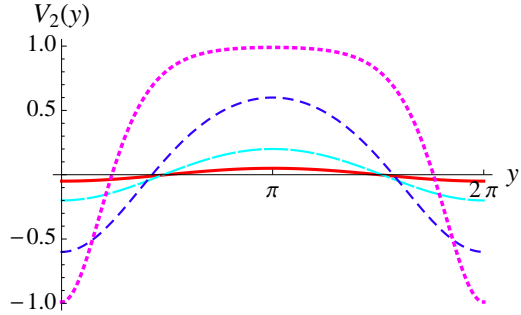}
\caption{Lam\'e potentials  $V_2$  for different values of $k$. As $k\to 0$ the intensity of the potential decreases due to 
	the $k^2$ term in (\ref{al}),  $V_2(y) \to 0$.  The unit for $y$
is taken $K/\pi$.}
\label{pot}
\end{figure}
Two independent solutions $\psi_{\pm}(y)$ of (\ref{al}) can be found in a closed form
\cite{Ganguly}
\begin{equation}\label{sol}
\psi_{\pm}(y)=\frac{{\cal H}(y\pm\alpha)}{\Theta(y)}e^{\mp y\zeta
(\alpha)}, \qquad \alpha\in \mathbb C,
\end{equation}
where ${\cal H}(y)$ and $\Theta(y)$ are theta functions and $\zeta(\alpha)$ is the Jacobi zeta function \cite{Abramovitz}. 
These theta functions satisfy $\Theta(y+2K)=\Theta(y)$ and ${\cal H}(y+2K)=-{\cal H}(y)$. The solutions (\ref{sol}) correspond 
to the eigenvalue $\tilde\epsilon^2$ which depends on the parameter $\alpha$ in the following manner,
\begin{equation}\label{energy}
 \tilde\epsilon^2=dn^2(\alpha,k).
\end{equation}
For the sake of convenience, we  divide the wave functions into a periodic part multiplied by a phase factor. Having in mind 
(anti-)periodicity of the theta functions, we can write
\begin{equation}\label{msols}
 \psi_{\pm}(y)=u_{\pm}(y)\,\exp\left(\mp i\, y\, p(\alpha)\right),\qquad u_{\pm}(y+2K)=u_{\pm}(y)
\end{equation}
where
\begin{equation}\label{p}
u_{\pm}(y)=\left(\frac{{\cal H}(y\pm\alpha)}{\Theta(y)}e^{\pm
i\frac{\pi y}{2K}}\right),\qquad { p(\alpha)}={-i\zeta
(\alpha)+\frac{\pi }{2K}}.
\end{equation}

The two independent (unnormalized) solutions $\Psi_\pm(y)$ for $H$ in (\ref{DiracLame2}) can be written with the help of 
$\mathcal{A}^{\dagger}$  (\ref{aas}) in the following manner 
\begin{equation}\label{eigenfunctions2}
\Psi_{\pm}(y)=\left(\begin{array}{c}\frac{i}{\tilde\epsilon}\,\mathcal{A}^{\dagger}\psi_{\pm}(y)
\\[1.5ex]
\psi_{\pm}(y)\end{array}\right)=\left(\begin{array}{c}
\frac{i}{\tilde\epsilon}(\mathcal{A}^{\dagger}\mp  p(\alpha))u_{\pm}(y)\\
 u_{\pm}(y) \end{array}\right)
 e^{\mp i y p(\alpha)}.
\end{equation}
Since differentiation of a function does not alter its periodicity, the spin-up component in (\ref{eigenfunctions2}) has 
the same phase factor as $\psi_{\pm}(y)$ and it has to be proportional to
$\psi_{\pm}(y+K)$, $\mathcal{A}^{\dagger}\psi_{\pm}(y)\sim\psi_{\pm}(y+K)$. For the special case $\tilde\epsilon=0$, 
the general solution is given by
\begin{equation}\label{zeromodes}
\Psi_0(y)=\left(\begin{array}{c}
\beta_2\, dn(y+K,k)\\[1.5ex]
\beta_1\, dn(y,k)\end{array}\right),\qquad
 H\Psi_0(y)=0,
\end{equation}
where $\beta_1$ and $\beta_2$ are arbitrary constants.

The admissible solutions $\Psi_{\pm}(y)$ and $\Psi_0(y)$ have to comply with the boundary condition (\ref{bc2}). 
Substituting (\ref{msols}) into (\ref{bc2}), we find that the values of the phase function $p(\alpha)$ in (\ref{eigenfunctions2}) 
has to satisfy the following equation 
\begin{equation}\label{pdelta}
 \pm p(\alpha)=\frac{\pi}{K}(\delta+l) ,\quad l\in\mathbb{Z}.
\end{equation}
We can see immediately that only the real values of $p(\alpha)$ are acceptable. Keeping in mind the definition (\ref{p}) 
of the phase factor $p(\alpha)$, the latter condition implies that only
the values of $\alpha$ such that $Re(\zeta(\alpha))=0$ are admissible. After some computations (see  Appendix 1) we find 
that the parameter $\alpha$ can be restricted to the following two intervals:
\begin{equation}\label{alfas}
\alpha=K+i\eta\quad {\rm or}\quad \alpha=i\eta, \quad  \quad  \eta\in[ 0,2K'].
\end{equation}
For these values of $\alpha$, the equation (\ref{pdelta}) takes the following form
\begin{equation}
\fl
\begin{array}{l}\label{bcelaborated1}
 p(\alpha)=\displaystyle
 -\zeta(\eta,k')+\frac{dn(\eta,k')sn(\eta,k')}{cn(\eta,k')}\\
\displaystyle \qquad\qquad   +s(\alpha)\frac{k^2sn(\eta,k')}{cn(\eta,k')dn(\eta,k')}
-\frac{\pi \eta}{2 K\,K'}+\frac{\pi}{2K}=\pm\frac{\pi}{K}(\delta+l)
\end{array}
\end{equation}
where $s(\alpha) = 0$ for $\alpha = i\eta$ and $s(\alpha) = 1$ for $\alpha = K+i\eta$. 

\begin{figure}[h!]
\centering
\includegraphics[scale=0.7]{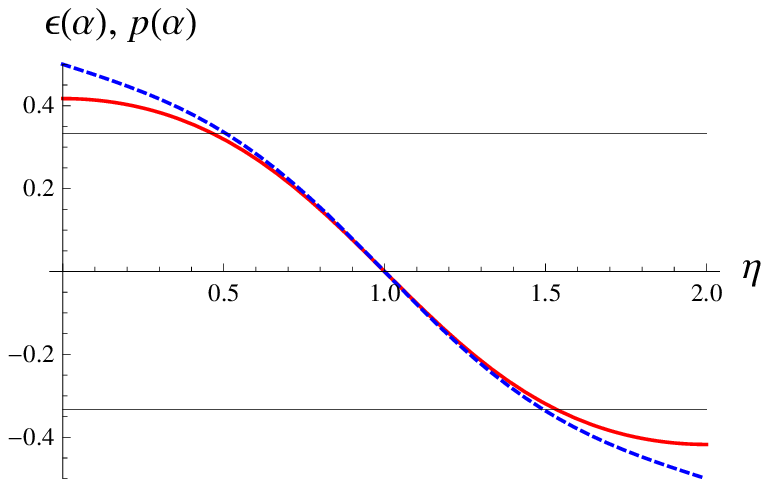}
\includegraphics[scale=0.7]{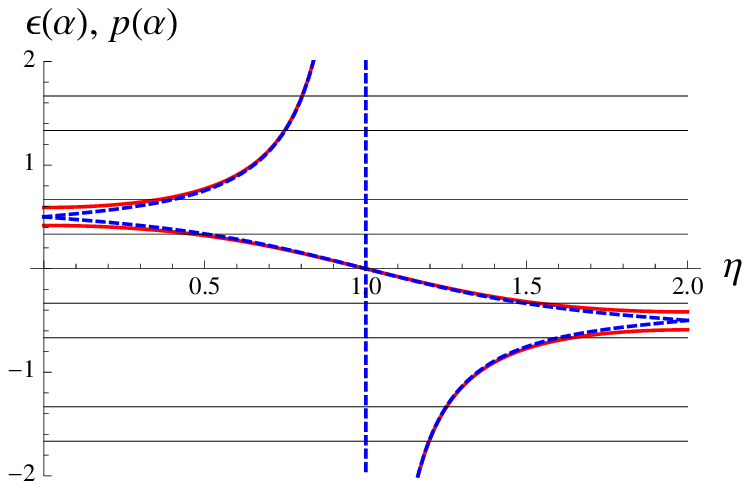}
\caption{ 
Left: The function $\frac{Kp(\alpha)}
{\pi}$ for $\alpha=K+i\eta$ (thin dashed line), $\epsilon(\alpha)$ (thick line)
and $\pm \delta$ (horizontal thin dashed line).
Right:
The functions $\frac{K\,p(\alpha)}{\pi}$ (dashing) and $\epsilon(\alpha)$ 
(continuous) as a function
of the parameter $\eta$ for the case $\alpha=K+i\eta$ (central band) and $\alpha=i\eta$
(upper and lower bands). The discrete values $\pm (\delta+l)$  are represented by thin  black lines for 
$l=0,\pm 1,\pm2,\pm3$. We have taken $k=\sqrt{0.5}$ and $\delta=\frac{1}{3} $. The unit of the horizontal axis is $K'$.}\label{fig_4ab}
\end{figure}

This transcendental equation must be solved numerically in order to find the admissible values of $\eta$, once $\delta$ 
and $l\in\mathbb Z$ are fixed. Some insight can be provided by a graphical solution of the equation, see Fig.\ \ref{fig_4ab} 
for illustration. The solutions of $\eta$ are given by the intersection of the horizontal lines which represent the r.h.s.\ 
of (\ref{bcelaborated1}) with the dashed curve of the function in the l.h.s.\ of (\ref{bcelaborated1}). For these solutions 
of $\eta$, the corresponding energy can be found on the solid curve.   

\begin{figure}[h!]
\centering
\includegraphics[scale=0.7]{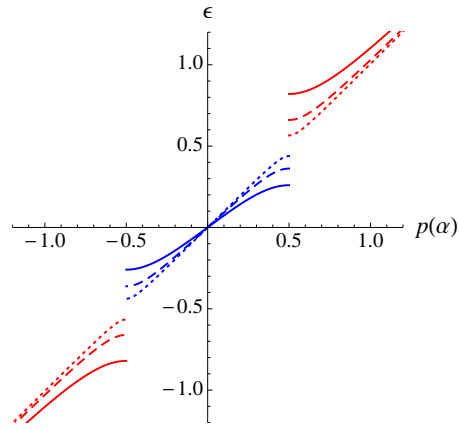}
\caption{Energies   as a function of the phase $p(\alpha)$
for different values of $k$. The continuous straight line is for the null field corresponding to the value $k=0$. }\label{fig_5}
\end{figure}

The central gap between positive and negative energies ${\epsilon}$ in the absence of transverse field ($A_{z}(\phi)=0$) is 
equal to $2\delta$,  which corresponds to the distance of the two straight lines in Fig.\ \ref{fig_4ab} (on the left side).  
The central gap of the system gets shrunk when the transverse magnetic field is switched on;  the corresponding positive 
(negative) energy curve lies below (above) the intersection of the phase curve with the horizontal line, see Fig.\ \ref{fig_4ab}. 
The positive and negative energies for different values of $k$  are represented in Fig.~\ref{fig_5}

\begin{figure}[h!]
\centering
\includegraphics[scale=0.7]{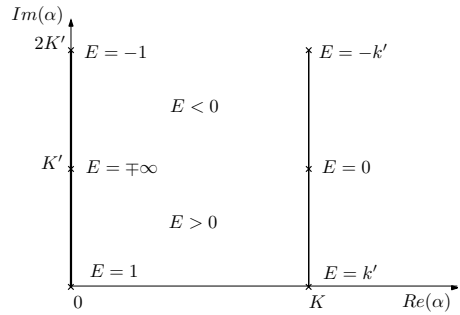}
\caption{Qualitative illustration of the dependence of energies $\tilde\epsilon(\alpha)$ on the complex values of~$\alpha$ }\label{fig_6}
\end{figure}

The energies are given by (\ref{energy}) in terms of $\alpha$. According to (\ref{alfas}) there are two types of admissible values 
of $\alpha$. For the type  $\alpha = K +i \eta$, we get the first energy band,
\begin{equation}\label{e1}
\tilde\epsilon(K+i\eta,k)) =
dn(K+i\eta,k)=k'\frac{cn(\eta,k')}{dn(\eta,k')},\qquad \eta\in [0,2K'].
\end{equation}
The second energy band is obtained with $\alpha = i \eta$,
\begin{equation}\label{e2}
\tilde\epsilon(i\eta,k)) =
dn(i\eta,k)=\frac{dn(\eta,k')}{cn(\eta,k')}, \qquad \eta\in [0,2K'].
\end{equation}

From the two formulas (\ref{e1}) and (\ref{e2}), we can get a qualitative insight into the behavior of the energy in dependence on 
the parameter $\alpha$. First, $\tilde\epsilon(\alpha)$ is positive for $\eta\in(0,K')$ while it acquires negative values for 
$\eta\in(K',2K')$. In the endpoints  of the intervals, we have $\tilde\epsilon(K)=k'$, $\tilde\epsilon(K+iK')=0$ and 
$\tilde\epsilon(K+2iK')=-k'$ while $\tilde\epsilon(0)=-\tilde\epsilon(2iK')=1$ and $\tilde\epsilon(iK'^{\mp})=\pm\infty$. 
See Fig.\ \ref{fig_6} for the illustration of these properties.

\section{Nonlinear supersymmetry of Dirac Hamiltonian}

Due to the specific form of the potential term, the Hamiltonian (\ref{DiracLame}) with integer $m$ belongs to the family of finite-gap 
operators that is associated with the stationary Ablowitz-Kaup-Newell-Segur (AKNS) hierarchy of nonlinear differential equations. 
The finite-gap systems possess an integral of motion $\mathcal{Y}$  given in terms of a higher-order differential operator. 
Its existence is deeply related with the theory of the integrable systems. It forms, together with the Hamiltonian, the celebrated 
Lax pair \cite{Gesztesy}. Its role in the description of both non-relativistic and relativistic periodic systems  was discussed 
extensively in the literature, see e.g. \cite{BdG,trisusy,PeriodicNanotubesII}.

In case of the Hamiltonian (\ref{DiracLame}), the symmetry operator acquires the following explicit form \cite{trisusy}
\begin{eqnarray}\label{Y}
  &&[H,\mathcal{ Y}]=0,\qquad \mathcal{ Y}=\left(\begin{array}{cc}0&Y(y)\\Y(y)^{\dagger}&0\end{array}\right),\end{eqnarray}
where
\begin{equation}
Y(y)=\frac{dn^{m+1}(y,k)}{cn^{2m+1}(y,k)}\left(\frac{cn^2(y,k)}{dn
(y,k)}\partial_y\right)^{2m}\frac{dn^{m-1}(y,k)}{cn^{2m-1}(y,k)}.
\end{equation}
The operator $\mathcal{Y}$ is uniquely determined by its kernel. It annihilates $2m$ formal eigenstates of $H$ corresponding to 
the eigenvalues $\tilde \epsilon_i$  ($i\in {1,\dots,2m}$). These states are antiperiodic and given in terms of Jacobi elliptic 
functions. The operator satisfies the following remarkable relation 
\begin{equation}\label{superalgebra}
\mathcal{ Y}^2=\prod_{i=1}^{2m}(H^2-\tilde \epsilon_i^2),
\end{equation}
where the right-hand side is identified as the spectral polynomial of the finite-gap system, see \cite{trisusy} for details.

The integral of motion of (\ref{DiracLame}) for $m=1$ has the following explicit form
\begin{eqnarray}
 \mathcal{ Y}&=&\sigma_1\left(\partial_y^2+\frac{k'^2}{2dn(y,k)^2}+\frac{1}{2}dn(y,k)^2\right)\nonumber\\
\fl &&
-ik^2\sigma_2\sqrt{\frac{sn(y,k)cn(y,k)}{dn(y,k)}}\partial_y\sqrt{\frac{sn(y,k)cn(y,k)}{dn(y,k)}}.
\end{eqnarray}
Its action on the wave functions (\ref{eigenfunctions2}) can be concluded directly from (\ref{superalgebra}) and (\ref{p}). 
As the operator $\mathcal{Y}$ is $2K$ periodic and because derivation of a function maintains its periodicity, we can write 
\begin{equation}
 \mathcal{ Y}\Psi_{\pm}(y)=
\tilde{u}_{\pm}e^{\mp ip(\alpha)y},\qquad \tilde{u}_{\pm}(y+2K)=\tilde{u}_{\pm}(y).
\end{equation}
Since $\mathcal{ Y}\Psi_{\tilde{\epsilon}}$ corresponds to the same energy as $\Psi_{\tilde{\epsilon}}$ (here, 
	$H\Psi_{\tilde \epsilon}={\tilde \epsilon}\Psi_{\tilde \epsilon}$) and satisfies the same boundary 
condition (it has the same phase factor), it must be proportional to $\Psi_{\tilde{\epsilon}}$. 
Having in mind (\ref{superalgebra}), we can conclude
\begin{equation}\label{Yaction}
 \mathcal{ Y}\Psi_{\tilde \epsilon}=\pm\sqrt{\prod_{i=1}^{2}({\tilde \epsilon}^2-{\tilde \epsilon}_i^2)}
 \Psi_{\tilde \epsilon}
 =\pm \sqrt{({\tilde \epsilon}^2-1)({\tilde \epsilon}^2-k'^2)}\Psi_{\tilde \epsilon}.
\end{equation}

Besides (\ref{Y}), the Hamiltonian (\ref{DiracLame}) (formally) commutes with the operator $\Gamma\equiv\sigma_3 \hat R$, 
where $\hat Ry\hat R=-y$ (there holds $cn(-y)=cn(y)$, $sn(-y)=-sn(y)$ and $dn(-y)=dn(y)$). However, the later operator 
does not preserve the boundary condition (\ref{bc2}) in general. To see it, we compute 
\begin{eqnarray}
\Gamma\Psi_{\pm}&=&\sigma_3\hat R\left(\begin{array}{c}
\frac{i}{\epsilon}(\mathcal{A}^{+}\mp  p(\alpha))u_{\pm}(y)\\
 u_{\pm}(y) \end{array}\right)
 e^{\mp i y p(\alpha)}\nonumber\\
&=&\sigma_3\left(\begin{array}{c}
\frac{i}{\epsilon}(\mathcal{A}^{+}\pm  p(\alpha))u_{\mp}(y)\\
-u_{\mp}(y) \end{array}\right)
e^{\pm i y p(\alpha)}=\Psi_{\mp}
\end{eqnarray}
where we used $ \hat Ru_{\pm}(y)=-u_{\mp}(y)$ and $\hat R\mathcal{A}^{\dagger}=-\mathcal{A}^{\dagger}\hat R$. 
The operator $\Gamma$ changes the sign of the phase factor, which collides with the boundary condition (\ref{bc2}) 
in general. The exception occurs when both $p(\alpha)=\frac{\pi}{K}(\delta+l)$ and $-p(\alpha)=\frac{\pi}{K}(\delta+\tilde{l})$  
for some integers $l$ and $\tilde{l}$. Then the change of sign in phase factor does not violate the boundary condition. 
This holds true for
\begin{equation}
 \delta=-\frac{l+\tilde{l}}{2}.
\end{equation}
Hence, $\Gamma$ is a good symmetry of the system provided that $\delta$ acquires (semi-)integer values. 

For $\delta=0$ or $\delta=1/2$, the operator $\Gamma$ commutes with $H$ but anticommutes with $\cal{Y}$. 
It suggests that the symmetries of the system can be conveniently treated in the framework of $N=2$ superalgebra, 
graded by  $\Gamma=\sigma_3\hat R$ and generated by the bosonic Hamiltonian (\ref{DiracLame}) and two fermionic supercharges, 
\begin{equation}\label{supercharge}
 Q_1={\cal Y},\quad Q_2=i\sigma_3\hat R{\cal Y},\quad [H,\Gamma]=\{Q_{a},\Gamma\}=0.
\end{equation}
Indeed, there hold the following relations
\begin{eqnarray}\label{nonsusy}
&&[H,Q_{a}]=0,\quad \{Q_a,Q_{b}\}=2\delta_{ab}\prod_{i=1}^{2m}(H^2-\tilde{\epsilon}_i^2),
\quad a,b=1,2.
\end{eqnarray}
Notice that the superalgebra differs from the standard supersymmetry as discussed in quantum mechanics 
\cite{cooper} where all the relations are linear in the generators. Here, the $N=2$ superalgebra 
(\ref{nonsusy}) is nonlinear as the anticommutator of the supercharges is a polynomial of second order 
in the Hamiltonian \cite{AIS}. Let us notice that this structure was discussed for periodic systems in 
\cite{BdG}. The supersymmetry of nonrelativistic systems based on nonlocal supercharges and graded by 
the parity operator was also discussed in \cite{trisusy}, \cite{Rsusy} or in \cite{NikitinRsusy}.

Below, we discuss the two specific configurations where the $N=2$ nonlinear supersymmetry exists, 
distinguished by the values of $\delta$:

\subsection{Metallic nanotubes: $\delta=0$}
The wave functions have to be periodic. It allows for existence of the ground states with zero energy.  
The explicit form of the two zero modes is
\begin{equation}\label{zerom}
 \Psi_{0\pm}=\left(\begin{array}{cc}
 dn(y,k)
 \\
 \pm dn(y+K,k)\end{array}\right). \quad 
\end{equation}
They satisfy the following relations, 
$$Q_a\Psi_{0\pm}=\pm k' \Psi_{0\pm},\quad H\Psi_{0\pm}=0,\quad a=1,2.$$
There is no central gap in the spectrum of the nanotubes. As discussed in  \cite{Novikov} the zero modes are 
protected by the standard supersymmetry presented in Sec. II, where the Hamiltonian (\ref{DiracLame2}) plays 
the role of supercharge.

\subsection{Maximally semi-conducting nanotubes: $\delta=1/2$}
The wave functions are required to be anti-periodic. In this regime, the nanotubes are maximally semi-conducting.  
The states $\Psi_{1\pm}$ corresponding to the threshold of the positive and negative energy spectrum, 
$\tilde{\epsilon}=\pm k'$, are annihilated by the supercharge $Q_1$ (and $Q_2$). Together with other two physical 
states $\Psi_{2\pm}$, corresponding to $\tilde{\epsilon}=\pm 1$, they form the kernel of $Q_1$. The explicit form 
of these states is   
\begin{equation}\label{nonlinstates}
 \Psi_{1\pm}=\left(\begin{array}{cc}\pm icn(y+K,k)\\ cn(y,k)\end{array}\right),\quad
\Psi_{2\pm}=\left(\begin{array}{cc}\pm i sn(y+K,k)\\ sn(y,k)\end{array}\right),
\end{equation}
where
\begin{equation}
\begin{array}{rl}
 H\Psi_{1\pm}=\pm k'\Psi_{1\pm},\qquad &\Psi_{1\pm}(y+2K) = - \Psi_{1\pm}(y)\\[1.5ex]
 \quad H\Psi_{2\pm}=\pm \Psi_{2\pm},\qquad  &\Psi_{2\pm}(y+2K) = - \Psi_{2\pm}(y) 
 \end{array}
\end{equation}
and 
$$Q_a\Psi_{b\pm}=0,\quad a,b\in\{1,2\}.$$


\section{Spectral stability in the $k_z\neq 0$ regime}

Up to now, we considered the system where the longitudinal momentum $k_z$ was vanishing. In the current section, 
we will extend the analysis to $k_z\neq0$ with the use of perturbation theory.

Let us consider the spectrum of the following operator
\begin{equation}
 {H_{k_z}}=H+k_z\sigma_2,
\end{equation}
where $H$ is the finite-gap Hamiltonian (\ref{DiracLame}). Let us suppose that $k_z$ is small enough to justify 
considering the term $\sigma_2k_z$ as a small perturbation. The first order correction $\Delta{\tilde \epsilon}$ 
to the energy $\tilde{\epsilon}$, where $\Psi_{{\tilde \epsilon}}=(\psi_{\tilde \epsilon}(y),\xi_{\tilde \epsilon}(y))^Te^{ip(\alpha)y}$ 
satisfies $(H-{\tilde \epsilon})\Psi_{{\tilde \epsilon}}=0$, is given by
\begin{eqnarray}\label{energycorrection}
\Delta{\tilde \epsilon}= k_z\int_{-K}^{K}\Psi_{{\tilde
\epsilon}}^{\dagger}\sigma_2\Psi_{\tilde \epsilon}d
y=2k_z\int_{-K}^{K} \mbox{Im}\left(\overline{\psi_{\tilde
\epsilon}(y)}\xi_{\tilde \epsilon}(y)\right)dy\nonumber\\
=-2k_z\int_{-K}^{K} \mbox{Im}\left(\beta\overline{\xi_{\tilde
\epsilon}(y)}\xi_{\tilde \epsilon}(y+K)\right)dy,
\end{eqnarray}
In the second line, we used the fact that the upper component $\psi_{\tilde{\epsilon}}$ is proportional to 
$\xi_{\tilde{\epsilon}}$ up to the shift of the coordinate, $\psi_{\tilde{\epsilon}}(y)=\beta\,\xi_{\tilde{\epsilon}}(y+K)$. 

The formula (\ref{energycorrection}) suggests that $\Delta{\tilde \epsilon}=0$ when both $\psi_{\tilde{\epsilon}}$ 
and $\xi_{\tilde{\epsilon}}$ are real. This happens for the zero mode (i.e. $\delta=0$); the stationary equation is 
decoupled and the upper and the lower component of the zero mode can be fixed as real functions.  

The robustness of energy levels can occur for nonzero levels as well, however, the specific properties of the system 
play more important role here.  Suppose we deal with antiperiodic boundary condition, i.e. $\delta=1/2$, which 
corresponds to the case of maximally semi-conducting nanotubes. Then the lower component $\xi_{\tilde\epsilon}(y)$ 
satisfies $\xi_{\tilde\epsilon}(y+2K)=-\xi_{\tilde\epsilon}(y)$. Next, we suppose that $\xi_{\epsilon}(y)$ is either 
even or odd with respect to the parity, i.e. $\xi_{\tilde\epsilon}(-y)=\tau \xi_{\tilde\epsilon}(y)$, where $\tau=\pm1$. 
Then we can write
\begin{eqnarray}\label{der1}
\int_{-K}^{K} \mbox{Im}\left(\beta\overline{\xi_{\tilde
\epsilon}(y)}\xi_{\tilde \epsilon}(y+K)\right)dy=-\int_{K}^{-K} \mbox{Im}\left(\beta\overline{\xi_{\tilde
\epsilon}(y)}\xi_{\tilde \epsilon}(y-K)\right)dy\nonumber\\=-\int^{K}_{-K} \mbox{Im}\left(\beta\overline{\xi_{\tilde
\epsilon}(y)}\xi_{\tilde \epsilon}(y+K)\right)dy,
\end{eqnarray}
where we made the substitution $y\rightarrow -y$ in the first step and then employed the parity and anti-periodicity of 
the wave functions. Since the left and the right hand side of the equality (\ref{der1}) differ just in sign, the integral 
has to be vanishing. Hence, the energy corresponding to the spinor with the required properties is robust with respect to 
small fluctuations of $k_z$. 
The states possessing required properties can be found in the kernel of the operator $\mathcal{Y}$, which is formed by $2m$ 
anti-periodic states $\Psi_{i}$, $i=1,...,2m$, by purely algebraic means, see \cite{trisusy}. 

Considering the specific case where $m=1$, one can prove much stronger statement: the first-order correction (\ref{energycorrection}) 
is vanishing for \textit{all} energy levels. Let us sketch the proof briefly, referring to Appendix 2 for more details. 
First, it is convenient to find explicitly the coefficient $\beta=\frac{\psi_{\tilde \epsilon}(y)}{\xi_{\tilde \epsilon}(y+K)}$. 
It is $\beta=i\,e^{\zeta(\alpha)K}$. Next, there holds the following relation  
\begin{equation}\label{odd}
\hat R(\overline{{\cal H}(y+\alpha)}{\cal H}(y+\alpha+K))=-{\cal H}(y+\alpha)\overline{{\cal H}(y+\alpha+K)}
\end{equation}
for $\alpha=i\eta$ and $\alpha=K+i\eta$. Substituting the explicit form of the wave function (\ref{p}) into the integral 
(\ref{energycorrection}), we get 
\begin{eqnarray}\label{ecorm}
\Delta\tilde\epsilon=-2k_z\int_{-K}^{K} \mbox{Im}\left(\beta\overline{\xi_{\tilde
\epsilon}(y)}\xi_{\tilde \epsilon}(y+K)\right)dy\nonumber\\
=-2k_z\int_{-K}^{K} Re\left(\frac{\overline{{\cal H}(y+\alpha)}{\cal H}(y+\alpha+K)}{\Theta(y)\Theta(y+K)}\right)dy=0.
\end{eqnarray}
Here we used the fact that $\zeta(\alpha)$ is purely imaginary and substituted the explicit form of $\beta$. The last equality 
is obtained since the integrand is an odd function of $y$, which follows from (\ref{odd}). The relation (\ref{ecorm}) manifests 
that the energy levels of $m=1$ case are robust with respect to small fluctuations of the longitudinal momentum $k_z$ for any 
value of $\delta$.

\section{Conclusions}
In the current paper, we have considered the continuum model of the single-wall  carbon nanotubes in the presence of a specific 
transverse magnetic field (\ref{approx})-(\ref{b0}). The model is quite flexible as it also allows to control the inhomogeneity 
of the field by the modular parameter $k$. For $k\rightarrow0$, the field tends to the homogenous one while for $k\rightarrow1$ 
it diverges from the constant value, see Fig. \ref{pot}. 

We elaborated in detail the case represented by the two-gap Hamiltonian (\ref{DiracLame2}). The wave functions and energies were 
found for the vanishing longitudinal momentum, $k_z=0$. The energies of the system were given as a solution of the transcendental 
equation (\ref{bcelaborated1}). We checked that the model manifests remarkable stability with respect to the fluctuations $k_z\neq 0$ 
of the longitudinal momentum. 

The considered systems belong to the family of finite-gap operators characteristic by existence of a nontrivial integral of motion 
(\ref{Y}). We showed that they give rise $N=2$ nonlinear supersymmetry (\ref{nonsusy}) for metallic and maximally semiconducting nanotubes.

In this article, we restricted ourselves to the dynamics in the vicinity of a single Dirac point. When both Dirac points are taken 
into account, the effective Hamiltonian is $4\times4$ block-diagonal operator. The $2\times2$ operators on its diagonal describe the 
system at each of the two Dirac points. In fact, the operator in (\ref{DiracLame2}) would correspond to one of them. Recently, 
the extended framework was used in the analysis of carbon nanotubes with finite-gap configurations of the pseudo-magnetic field 
generated by the axial twist \cite{PeriodicNanotubesII}. It was observed that when both Dirac points are taken into account, 
there emerges a $so(3)\oplus u(1)$ algebra of integrals of motion for the system with time-reversal symmetry. The question 
arises whether a similar algebraic structure could be detected in the current system when both Dirac points would be considered. 
Despite that our current setting lacks time-reversal symmetry (it is violated by the external magnetic field), we suppose that 
the algebra could exist for 
any strength of the parallel magnetic field $B_{\parallel}$ as long as $\omega=0$. 

The treatment here presented opens a number of interesting questions. 
As the wave functions of the Lam\'e Hamiltonian are known explicitly for any integer $m$, it suggests that it could be feasible 
to extend our results to the  systems with a generic, integer valued, coupling constant $m$.  In particular, we have in mind the  
existence of the hidden $N=2$ nonlinear supersymmetry or the spectral stability of the system. Our present results based on finite-gap 
Hamiltonian provide a good basis for construction of other solvable configurations of external magnetic field. We have in mind to apply 
Darboux transformations \cite{DiracDarboux} in the construction of new solvable Hamiltonians from the systems presented in this paper. 
Let us notice in this context that this technique proved to be very useful in the analysis of twisted carbon nanotubes \cite{PeriodicNanotubesI}.  
The results supplied along this paper system can also serve as a test field for the analysis of a wider class of configurations with the 
use of rigorous methods like those applied recently in the study of low-dimensional nonrelativistic systems with external magnetic field \cite{Krejcirik}.
However, these considerations go beyond the scope of the present work.

\section*{Appendix 1: Computation of the phase of the eigenfunctions}

Using the appropriate formulas for the
Jacobi zeta functions \cite{Abramovitz}, we can find
\begin{eqnarray}\label{zeta}
 \zeta(\alpha,k)&=&\zeta(Re(\alpha),k)+k^2\frac{s_I^2\,s\,c\,d}{c_I^2+k^2s_I^2s^2}
 \\&&+i\left(-\zeta(Im(\alpha),k')-\frac{\pi Im(\alpha)}{2K(k)K(k')}+\frac{d_Is_I}{c_I}\right)
 \\&&-ik^2\frac{s^2 s_I\, d_I}{c_I(c_I^2+k^2s^2s^2_I)},
\end{eqnarray}
where we abbreviated
\begin{eqnarray}
 &&s=sn(Re(\alpha),k),\quad c=cn(Re(\alpha),k),\quad d=dn(Re(\alpha),k),\\
 &&s_I=sn(Im(\alpha),k'),\quad c_I=cn(Im(\alpha),k'),\quad d_I=dn(Im(\alpha),k').
\end{eqnarray}
In order to have $p(\alpha)$ real valued (see (\ref{p})), the real part of (\ref{zeta}) has to be eliminated. It can be done by 
fixing $Re(\alpha)$ as an integer multiple of $K$, i.e.
\begin{equation}
 \alpha=n K+i\eta,\quad \eta\in\mathbb{R},\quad n\in\mathbb{Z}.
\end{equation}
Then there holds $\zeta(l K,k)=0$ and $s\,c=0$ and the real part of (\ref{zeta}) vanishing. We can write
\begin{eqnarray}\label{pp}
 p(\alpha)&=&-\zeta(\eta,k')+\frac{dn(\eta,k')sn(\eta,k')}{cn(\eta,k')}+s(n)
 \frac{k^2sn(\eta,k')}{cn(\eta,k')dn(\eta,k')}\nonumber\\
&&-\frac{\pi \eta}{2KK'}+\frac{\pi }{2K},\quad
\end{eqnarray}
where $s(n)=\frac{(1-(-1)^n)}{2}$.
The relation (\ref{p}) together with (\ref{pdelta}) determine the relevant values of $\alpha$. The physically relevant values of 
$\alpha$ (or $\eta$) are given by the following equation
\begin{eqnarray}\label{bcelaborated}
 p(\alpha)&=&-\zeta(\eta,k')+\frac{dn(\eta,k')sn(\eta,k')}{cn(\eta,k')}+\epsilon(n)\frac{k^2sn(\eta,k')}{cn(\eta,k')dn(\eta,k')}\nonumber\\
&&-\frac{\pi \eta|_{mod(2K')}}{2KK'}+\frac{\pi}{2K}\left(1-2\left[\frac{\eta}{2K'}\right]\right)\\&=&\pm \frac{\pi}{K}(\delta+l),\quad
\end{eqnarray}
where $[\ .\ ]$ denotes the integer part of a number and $\eta=\eta|_{mod(2K')}+2K'[\frac{\eta}{2K'}]$.
Without loss of generality, we can fix $\alpha=K+i\eta$ or $\alpha=i\eta$ with $\eta\in\langle0,2K'\rangle$.\\

\section*{Appendix 2: Stability of energy levels}
The function ${\cal H}(y)$ is $2K$ antiperiodic and odd with respect to the parity operator $\hat R$, ${\cal H}(y+2K)=-{\cal H}(y)$, 
$\hat R{\cal H}(y)={\cal H}(-y)=-{\cal H}(y)$. The eta function is $2K$ periodic and even with respect to the parity, 
$\Theta(y+2K)=-\Theta(y)$, $\hat R\Theta(y)=\Theta(-y)=\Theta(y)$. Next, there holds $\overline{{\cal H}(y+\alpha)}={\cal H}(y+\overline{\alpha})$ 
provided that $y$ is real. These properties can be obtained directly from the definition, see \cite{Abramovitz}.

There holds the following relation between the up and down components of the eigenfunction $\Psi_{\tilde{\epsilon}}=(\psi,\xi)^T$,
\begin{equation}
 \psi(y)=\beta\,\xi(y+K),
\end{equation}
where $\beta$ is a complex number that should be specified. There also holds 
\begin{equation}
 \frac{i}{\tilde{\epsilon}}\mathcal{A}^{\dagger}\xi=\psi,\quad \frac{1}{{\tilde\epsilon}^2}\mathcal{A}\mathcal{A}^{\dagger}\xi=\xi,
 \quad  \mathcal{A}^{\dagger}(y+K)=-\mathcal{A}(y).
\end{equation}
Then we can write
\begin{eqnarray}
 \xi(y)&=&\frac{1}{\tilde\epsilon^2}\mathcal{A}\mathcal{A}^{\dagger}\xi(y)=-\frac{i}{\tilde\epsilon}\mathcal{A}\psi(y)
 =-\frac{i}{\tilde\epsilon}\mathcal{A}\beta\xi(y+K)\nonumber\\
 &=&\frac{i}{\tilde\epsilon}\mathcal{A}^{\dagger}(y+K)\beta\xi(y+K)=\beta^2\xi(y+2K)
\end{eqnarray}
so that $\beta^2=\frac{\xi(y)}{\xi(y+2K)}$. Keeping in mind that $\xi=\frac{{\cal H}(y+\alpha)}{\Theta(y)}e^{-\zeta(\alpha)y}$, 
we can write
\begin{eqnarray}
\fl
 \beta^2=\frac{\xi(y)}{\xi(y+2K)}
 =\frac{{\cal H}(y+\alpha)\Theta(y+2K)e^{-\zeta(\alpha)y}}{{\cal H}(y+\alpha+2K)\Theta(y)e^{-\zeta(\alpha)(y+2K)}}=-e^{2\zeta(\alpha)K}.\nonumber\\
\end{eqnarray}
Hence, we get $\beta=i e^{\zeta(\alpha)K}$. 

Now, we can simplify the integrand $I$ of the first-order energy correction,
\begin{eqnarray}
 I=(\overline{\psi},\overline{\xi})\sigma_2\left(\begin{array}{c}\psi\\\xi\end{array}\right)
 = i\beta\overline{\xi(y)}\xi(y+K)-i\overline{\beta\xi(y+K)}\xi(y)\nonumber\\=-Im(\beta\,\overline{\xi(y)}\,\xi(y+K)),
\end{eqnarray}
by inserting explicit form of $\xi$ and substituting the explicit value for $\beta$. We get
\begin{eqnarray}
 I=Im\left(i\, e^{\zeta(\alpha)K}\frac{\overline{{\cal H}(y+\alpha)}{\cal H}(y+\alpha+K)}
 {\Theta(y)\Theta(y+K)}e^{-\overline{\zeta(\alpha)}y}e^{-\zeta(\alpha)(y+K)}\right)\nonumber\\
=Im\left(i\,\frac{\overline{{\cal H}(y+\alpha)}{\cal H}(y+\alpha+K)}{\Theta(y)\Theta(y+K)}\right)
=Re\left(\,\frac{\overline{{\cal H}(y+\alpha)}{\cal H}(y+\alpha+K)}{\Theta(y)\Theta(y+K)}\right).\nonumber
\end{eqnarray}
Now, we will show that the integrand $I$ is odd with respect to the parity. Since $\Theta(y)$ is $2K$ periodic 
and even with respect to the parity, it is sufficient to show that 
$$
\hat R(\overline{{\cal H}(y+\alpha)}{\cal H}(y+\alpha+K))=-{\cal H}(y+\alpha)\overline{{\cal H}(y+\alpha+K)}.
$$
Taking $\alpha=K+i\eta$, we can write
\begin{eqnarray}\fl
 \hat R(\overline{{\cal H}(y+{\alpha})}{\cal H}(y+\alpha+K))&=&{\cal H}(-y+\overline{\alpha}){\cal H}(-y+\alpha+K)\nonumber\\
&=&{\cal H}(-y+K-i\eta){\cal H}(-y+i\eta+2K)\nonumber\\
 &=&-{\cal H}(-y-K-i\eta){\cal H}(-y+i\eta+2K)\nonumber\\&=&{\cal H}(y+K+i\eta){\cal H}(-y+i\eta-2K)\nonumber\\
&=&-{\cal H}(y+\alpha){\cal H}(y-i\eta+2K)=-{\cal H}(y+\alpha){\cal H}(y+\overline{\alpha}+K)\nonumber\\
&=&-{\cal H}(y+\alpha)\overline{{\cal H}(y+\alpha+K)}
\end{eqnarray}
Similar result can be obtained for $\alpha=i\eta$.
We showed that the integrand $I$ is an odd function. Hence, the first energy correction is vanishing identically,
\begin{equation}
 \int_{-K}^{K}(\overline{\psi},\overline{\xi})\sigma_2
 \left(\begin{array}{c}\psi\\
 \xi\end{array}\right) dy=0.
\end{equation}

\section*{Acknowledgments}
This work is partially supported by the Spanish MEC (FIS2009-09002)
and by the GA\v CR Grant P203/11/P038 of the Czech Republic.


\end{document}